\documentclass[11pt,b5paper,twoside]{article}
\usepackage{latexsym,amssymb,amsmath}
\usepackage{tikz,pgf}

\newtheorem{theo}{Theorem}
\newtheorem{defi}[theo]{Definition}

\newtheorem{lemm}[theo]{Lemma}

\newenvironment{proof}{\bizo}{\kocka}\def\bizo{\par\noindent{\textbf{Proof.\ }}}
\def\kocka{\nobreak\hspace*{\fill}\nobreak$\Box$\par\smallskip}

\newcommand{\key}[1]{\textbf{#1}}
\newcommand{\tsc}[1]{\textrm{\textsc{#1}}}

\newcommand{\alge}{%
\begin{tabbing}%
99 \= xxx\=xxx\=xxx\=xxx\=xxx\=xxx\=xxx\=xxx\=xxx\=xxx\=xxx \+ \kill 
}

\newcommand{\algeN}{%
\begin{tabbing}%
\hspace*{1pt}999\=xxx\=xxx\=xxx\=xxx\=xxx\=xxx\=xxx\=xxx\=xxx\=xxx\=xxx \+ \kill
}

\newcommand{\algv}{%
\end{tabbing}
} 

\newenvironment{algN}[1]{%
\vspace{4mm}  
\vbox\bgroup\noindent\tsc{#1}%
\vspace*{-2mm}  
\algeN}
{
\algv\egroup
\vspace{0mm}  
} 

\begin{document}

\begin{center}{\Large\bf
Super-$d$-complexity of finite words
}
\end{center}
\begin{center}{\large\bf\noindent
Zolt\'an K\'asa}\\[2mm]

Sapientia Hungarian University of Transylvania\\Cluj--M. Ciuc--Tg. Mure\c s

Department of Mathematics and Informatics, Tg. Mure\c s/Marosv\'as\'arhely
\\[1mm]\texttt{
kasa@ms.sapientia.ro}
\end{center}
\vspace*{7mm}

\begin{abstract} In this paper we introduce and study a new complexity measure for finite words. For positive integer $d$ special scattered subwords, called super-$d$-subwords, in which the gaps are of length at least $(d-1)$, are defined.  We give methods to compute super-$d$-complexity (the total number of different super-$d$-subwords) in the case of rainbow words (with pairwise different letters) by recursive algorithms, by mahematical formulas and by graph algorithms. In the case of general words, with letters from a given alphabet without any restriction, the problem of the maximum value of the super-$d$-complexity of all words of length $n$ is presented.
\end{abstract}


\section{A new complexity measure: the super-$d$-complexity} 
Sequences of characters called \emph{words} or \emph{strings} are widely studied in combinatorics, and used in  
various fields of sciences (e.g. chemistry, physics, social sciences, biology \cite{KZ:fiz,KZ:elzinga,KZ:elzinga2,KZ:bio} etc.).
The elements of a word  are called \emph{letters}. A contiguous part of a word (obtained by erasing a prefix or/and a suffix) is a \emph{subword} or \emph{factor}. If we erase arbitrary letters from a word, what is obtained is a \emph{scattered subword}. Special scattered subwords, in which the consecutive letters are at distance at most $d$ $(d\ge 1)$ in the original word, are called $d$-\emph{subwords} \cite{KZ:ivanyi, KZ:kasa}. In this paper we define another kind of scattered subwords, in which the original distance between two letters which are consecutive in the subword, is at least $d$ $(d\ge 1)$, these will be called \emph{super-$d$-subwords}.  

One can easily observe that in any given word, the 1-subwords are exactly the (ordinary) subwords, and the
super-1-subwords are exactly the scattered subwords.
            
The \emph{complexity of a word} is defined as the total number of its different subwords. The definitions of \emph{$d$-complexity} and \emph{super-$d$-complexity} are similar.

For a (finite) alphabet  $\Sigma$, as usual,  $\Sigma^n$  and
          $\Sigma^*$  are the sets of all words of length $n$, and of all finite
          words, respectively, over  $\Sigma$.

In order to formalize the above, we introduce the following two definitions.

\begin{defi} 
Let  $n$, $d$ and $s$ be positive integers, and $u=x_1x_2\ldots x_n\in \Sigma^n$. A \textbf{super-$d$-subword} of length $s$ of $u$ is defined as $v=x_{i_1}x_{i_2}\ldots x_{i_s}$ where 

$i_1\ge 1$, 

$d\le i_{j+1}-i_j< n$  for $j=1,2,\ldots, s-1$,

$i_s\le n.$
\end{defi}

\begin{defi} 
The \textbf{super-$d$-complexity} of a word is the total number of its different super-$d$-subwords.  
\end{defi}

The super-$2$-subwords of the word \emph{abcdef}  are the following: \emph{a, ac, ad, ae, af, ace, acf, adf, b, bd, be, bf, bdf, c, ce, cf, d,  df, e, f}, therefore the super-2-complexity of this word is $20.$

\section{Computing the super-$d$-complexity of rainbow words}
Words with pairwise different letters are called \emph{rainbow words}.
The super-$d$-complexity of a rainbow word of length $n$ does not depends on what letters it contains, and is denoted by $S(n,d)$.
            
Let us denote by $b_{n,d}(i)$ the number of super-$d$-subwords which begin at the $i$-th position  in a rainbow word of length $n$. Using our previous example (\emph{abcdef}), we can see that $b_{6,2}(1)=8$, $b_{6,2}(2)=5$, $b_{6,2}(3)=3$, $b_{6,2}(4)=2$, $b_{6,2}(5)=1$,  and $b_{6,2}(6)=1$. 

We immediately get the following formula:
\begin{equation}
b_{n,d}(i)= 1 + b_{n,d}(i\!+\!d)+ b_{n,d}(i\!+\!d\!+\!1)+\!\cdots\! +  b_{n,d}(n),   \label{KZ:1}
\end{equation}

\noindent for $n> d, 1\le i\le n-d$, and

\[b_{n,d}(1)=1  \textrm{ for  } n\le d.\]

\noindent The super-$d$-complexity of rainbow words can be computed by the following formula:
\begin{equation}
S(n,d)= \sum_{i=1}^{n}{b_{n,d}(i)}. \label{KZ:ketto}
\end{equation}

\noindent This can be expressed also as
\begin{equation} S(n,d)=\sum_{k=1}^{n}{b_{k,d}(1)}, \label{KZ:t1} \end{equation}
because of the formula
\[S(n+1,d)=S(n,d)+b_{n+1,d}(1).\]

\begin{table}
\begin{center}
\begin{tabular}{|r|rrrrrrrrrrr|}\hline
\begin{picture}(15,10)\put(-6,10){\line(2,-1){27}}\put(-2,-2){$n$}\put(13,3){$d$}\end{picture}& 1& 2  & 3  & 4  & 5 &6 &7 &8 &9&10&11\\ \hline
1& 1 &1&1&1&1&1&1&1&1&1&1 \\
2& 3 &2&2&2&2&2&2&2&2&2&2 \\
3& 7& 4&3&3&3&3&3&3&3&3&3\\
4& 15&7 &5&4&4&4&4&4&4&4&4\\
5& 31&12 &8&6&5&5&5&5&5&5&5\\
6 & 63& 20  &12 &9 & 7&6&6&6&6&6&6\\
7 & 127&33  &18 &13 &10 &8 &7&7&7&7&7\\
8 & 255&54  &27& 18 & 14 & 11 &9&8&8&8&8\\
9 & 511&88  &40 &25 & 19&15&12&10&9&9&9\\
10 & 1023& 143&59 &35 & 25&20&16&13&11&10&10\\
11 &2047& 232&87 &49 & 33&26&21&17&14&12&11\\
12 &4095& 376&128 &68 & 44&33&27&22&18&15&13\\
\hline
\end{tabular}\caption{Values of $S(n,d)$.}
\end{center}
\end{table}

\bigskip
In the case $d=1$ the complexity  $S(n,1)$ can be computed easily: $S(n,1)= 2^n-1$. This is equal to  the $n$-complexity of  rainbow words of length $n$.

In the sequel we will present different methods to compute the super-$d$-complexity of the rainbow words. In the description of algorithms  the pseudocode conventions from \cite{KZ:CLRS} are used.

\subsection{Computing by recursive algorithm}
 From (\ref{KZ:1}) for the computation of $b_{n,d}(i)$ the following algorithm is obtained. The numbers $b_{n,d}(k) \,   (k=1,2,\ldots)$ for a given $n$ and $d$ are obtained in  the array $b=(b_1,b_2, \ldots $), which is a global parameter in the following algorithms. Initially all these elements are equal to $-1$.
The call for the given $n$ and $d$ and the desired $i$ is:

\begin{tabbing}
\key{Input} $n,d,i$\\
\key{for} \= $k\leftarrow 1$ \key{to} $n$\\
          \> \key{do} $b_k\leftarrow -1$\\
\textsc{B}$(n,d,i)$ \\         
\key{Output} $b_1, b_2, \ldots, b_n$
\end{tabbing}

\medskip
The recursive algorithm is the following:

\begin{algN}{B($n,d,i$)}
1 \' $p\leftarrow 1$ \\
2 \' \key{for} \= $k\leftarrow i+d$ \key{to} $n$ \\
3 \'           \> \key{do} \= \key{if}  \= $b_k=-1$\\
4 \'           \>          \>     \>  \key{then} \textsc{B}$(n,d,k)$\\
5 \'           \>          \> $p\leftarrow p+b_k$              \\
6 \' $b_i\leftarrow p$ \\ 
7 \'  \key{return}
\end{algN}

If the call is $B(8,2,1)$, the elements will be obtained in the following order:
$b_7=1$, $b_8=1$, $b_5=3$, $b_6=2$, $b_3=8$, $b_4=5$, and $b_1=21$.

\bigskip
\begin{lemm}\label{KZ:l1} $b_{n,2}(1)=F_{n},$ where $F_n$ is the $n$-th Fibonacci number.
\end{lemm}
\begin{proof}
Let us consider a rainbow word $a_1a_2\ldots a_n$ and let us count all of its super-$2$-subwords which begin with $a_2$. If we change $a_2$ for $a_1$ in each super-$2$-subword which begin with $a_2$, we again obtain super-$2$-subwords. If we prefix an $a_1$ to each super-$d$-subword which begin with $a_3$, we again obtain super-$d$-subwords.  Thus
\[b_{n,2}(1)=b_{n-1,2}(1)+b_{n-2,2}(1).\]
So $b_{n,2}(1)$ is a Fibonacci number, and because $b_{1,2}(1)=1$, we obtain $b_{n,2}(1)=F_n.$
\end{proof}

\begin{theo}\label{KZ:tt}
$S(n,2)=F_{n+2}-1$, where $F_n$ is the $n$-th Fibonacci number.
\end{theo}
\begin{proof}
From (\ref{KZ:t1}) and Lemma \ref{KZ:l1}:
\begin{eqnarray*}
S(n,2)&=&b_{1,2}(1)+ b_{2,2}(1)+ b_{3,2}(1)+b_{4,2}(1)+ \cdots + b_{n,2}(1)\\
      &=& F_1+F_2+\cdots + F_{n}\\
      &=& F_{n+2}-1. 
\end{eqnarray*}

\vspace*{-0.8cm}
\end{proof}

\smallskip
Introducing the notation $M_{n,d}=b_{n,d}(1)$, then by the formula
\[b_{n,d}(1)=b_{n-1,d}(1)+b_{n-d,d}(1),\]
a generalized middle sequence (see the sequence A000930\footnote{From \cite{KZ:online}: $a_0=a_1=a_2=1$; thereafter $a_n=a_{n-1} + a_{n-3}$. Might be called the Middle Sequence, since it is a cross between the Fibonacci sequence (A000045) and the Padovan sequence (A000931).} in \cite{KZ:online}) will be obtained in the following, recursive way:
\begin{eqnarray}
M_{n,d}\!\!\!\!&=&\!\!\!\!M_{n-1,d} +M_{n-d,d}, \quad\textrm{for } n\ge d\ge 2,\label{KZ:MMM}\\
M_{0,d}\!\!\!\!&=&\!\!\!\!0, \, M_{1,d}=1,\, \ldots,\, M_{d-1,d}=1.\nonumber
\end{eqnarray}
Let us call this sequence \emph{$d$-middle sequence}. Because of the  equality $M_{n,2}=F_n$, the $d$-middle sequence can be considered as a generalization of the Fibonacci sequence.

The $d$-middle sequence defined in (\ref{KZ:MMM}) is a little different from  the generalization of the sequence A000930 in \cite{KZ:online} because of its initial values.

The next algorithm  computes $M_{n,d}$, by using an array $M_0, M_1, \ldots, M_{d-1}$ to store the necessary previous elements:

\begin{algN}{Middle($n,d$)}
1 \' $M_0 \leftarrow 0$ \\
2 \' \key{for} \= $i \leftarrow 1$ \key{to} $d-1$ \\
3 \'           \> \key{do} $M_i\leftarrow 1 $\\
4 \'  \key{for} \= $i \leftarrow d$ \key{to} $n$ \\ 
5 \'    \> \key{do} \= $M_{i \textrm{ \scriptsize mod } d} \leftarrow M_{(i-1) \textrm{ \scriptsize mod } d} + M_{(i-d) \textrm{ \scriptsize mod } d} $  \\
6 \'    \>          \> print  $M_{i \textrm{ \scriptsize mod } d} $                 \\ 
7 \'  \key{return}
\end{algN}

\medskip
Using the generating function $M_d(z)=\displaystyle\sum_{n\ge 0 }{M_{n,d}z^n}$,  the following closed formula is obtained:
\begin{equation}
M_d(z)=\frac{z}{1-z-z^d}. \label{KZ:gen}
\end{equation}
This can be used to compute the sum $s_{n,d}=\displaystyle\sum_{n=1 }^{n}{M_{i,d}}$, which is the coefficient of $z^{n+d}$ in the expansion of the function 
\[\frac{z^{d}}{1-z-z^d}\cdot\frac{1}{1-z}= \frac{z^{d}}{1-z-z^d}+\frac{z}{1-z-z^d}-\frac{z}{1-z}.\]
So $s_{n.d}=M_{n+(d-1),d}+M_{n,d}-1 = M_{n+d,d}-1.$ Therefore
\begin{equation} 
\sum_{i=1}^{n}{M_{i,d}}= M_{n+d,d} -1. \label{KZ:m}
\end{equation}

\begin{theo}\label{KZ:t2}
$S(n,d)=M_{n+d,d}-1$, where $n>d$ and $M_{n,d}$ is the $n$-th element of the $d$-middle sequence.
\end{theo}

\begin{proof} The proof is similar to that in Theorem \ref{KZ:tt} taking into account formula (\ref{KZ:m}).
\end{proof}

\subsection{Computing by mathematical formulas}
\begin{theo}  $S(n,d)=\displaystyle\sum_{k\ge 0}{\binom{n-(d-1)k}{{k+1}}}$, \ for $n\ge 2, d\ge 1$.
\end{theo}
\begin{proof}
   Let us consider the generating function $\displaystyle G(z)=\frac{1}{1-z}= 1+z+z^2+\cdots  $. Then, taking into account the formula (\ref{KZ:gen}) we obtain $M_d(z)=zG(z+z^d)=z + z(z+z^{d})+z(z+z^{d})^2+\cdots +z(z+z^{d})^i +\cdots .$ The general term in this expansion is equal to  
   \[ z^{i+1}\sum_{k=1}^{i}{\binom{i}{k}z^{(d-1)k}}, \]
and the coefficient of $z^{n+1}$ is equal to \[ \sum_{k\ge 0}^{}{\binom{n-(d-1)k}{k} }  .\]  The coeeficient of $z^{n+d}$ is 
\begin{equation}
   M_{n+d,d}= \sum_{k\ge 0}^{}{\binom{n+d-1 -(d-1)k}{k} }. \label{KZ:Mnd}
\end{equation} 
By Theorem \ref{KZ:t2}  $S(n,d)=M_{n+d,d}-1$, and an easy computation yields 
   \[S(n,d)= \sum_{k\ge 0}{{n-(d-1)k} \choose {k+1}}.\]   

\vspace*{-1.5\baselineskip}
\end{proof}

\begin{theo}  $b_{n+1,d}(1)=\displaystyle\sum_{k\ge 0}{{n-(d-1)k} \choose {k}}$,  for $n\ge 1, d\ge 1$.
\end{theo}
\begin{proof}
   From $b_{n+1,d}(1)=M_{n+1,d}$ and (\ref{KZ:Mnd}):
   \[b_{n+1,d}= \sum_{k\ge 0}^{}{\binom{n -(d-1)k}{k} }. \]

\vspace*{-1.5\baselineskip}
\end{proof}

\subsection{Computing by graph algorithms}
To compute the super-$d$-complexity of a rainbow word of length $n$, let us consider the word  $a_1a_2\ldots a_n$ and the correspondig digraph $G=(V, E)$, with 

$V=\big\{ a_1, a_2, \ldots, a_n \big\}$, 

$E=\big\{ (a_i,a_j) \mid j-i\ge d, \, i=1,2,\ldots, n,  j=1,2,\ldots, n  \big\}$. 

\begin{figure}[t]
\begin{center}
\begin{tikzpicture}
\begin{scope}[>=latex]
\filldraw[]     (1,2) circle (2pt) 
                (2,2) circle (2pt)
                (3,2) circle (2pt)  
                (4,2) circle (2pt)
                (5,2) circle (2pt) 
                (6,2) circle (2pt);
\draw [->] (1,2) .. controls (1.5,2.5) and (2.5,2.5) .. (2.95,2.05);
\draw [->] (1,2) .. controls (2,2.8) and (3,2.8) .. (3.95,2.05);
\draw [->] (1,2) .. controls (2,3.1) and (4,3.1) .. (4.95,2.05);
\draw [->] (1,2) .. controls (2,3.4) and (5,3.4) .. (5.95,2.05);
\draw [->] (3,2) .. controls (3.5,2.5) and (4.5,2.5) .. (4.95,2.05);
\draw [->] (3,2) .. controls (4,2.8) and (5,2.8) .. (5.95,2.05);
\draw [->] (4,2) .. controls (4.5,1.4) and (5.5,1.4) .. (5.95,1.95);
\draw [->] (2,2) .. controls (2.5,1.4) and (3.5,1.4) .. (3.95,1.95);
\draw [->] (2,2) .. controls (3,0.9) and (4,0.9) .. (4.95,1.95);
\draw [->] (2,2) .. controls (2.5,0.6) and (5.5,0.6) .. (5.95,1.95);
\node (1) at (1,1.75) {$a$};
\node (2) at (1.95,1.75) {$b$};
\node (3) at (3,1.75) {$c$};
\node (4) at (4,1.75) {$d$};
\node (5) at (5,1.75) {$e$};
\node (6) at (6.05,1.75) {$f$};
\end{scope}
\end{tikzpicture}
\end{center}\vspace*{-0.8cm}
\caption{Graph for $2$-subwords when $n=6.$}\label{KZ:fig1}
\end{figure}
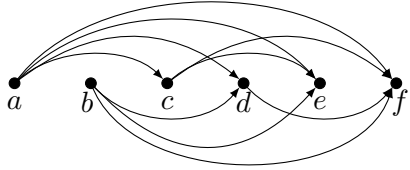

For $n=6, d=2$ see Figure \ref{KZ:fig1}.

The adjacency matrix  $A=\big(a_{ij}\big)_{\!\!\!\tiny\begin{array}{c}i\!\!=\!\!\overline{1,\!n}\\
                                                            j\!\!=\!\!\overline{1,\!n}\end{array}}$ 
of the graph is defined by:

\[a_{ij}=\left\{  \begin{array}{ll}
                    1,  &  \textrm{if } j-i\ge d,\\
                    0,     &  \textrm{otherwise},
                 \end{array}  \quad \textrm{  for  }   i=1,2,\ldots, n, j=1,2,\ldots, n.                        
           \right.
\]

Because the graph has no directed cycles, the entry in row $i$ and column $j$ in  $A^k$ (where $A^k=A^{k-1}A$, with $A^1=A$) will represent the number of $k$-length  directed paths from $a_i$ to $a_j$.  If $I$ is the identity matrix  (with elements equal to 1 only on the first diagonal, and 0 otherwise), let us define the matrix $R= (r_{ij})$:
\[R= I+A+A^2+\cdots + A^k, \textrm{ where } A^{k+1}=O \, (\textrm{the null matrix}).\]
The super-$d$-complexity of a rainbow word is then
\[ S(n,d)= \sum_{i=1}^{n}{\sum_{j=1}^{n}{r_{ij}}}.\]
To compute matrix $R$, we define a variant of the well-known Warshall algorithm (for this  see for example \cite{KZ:baase}): 
 
\begin{algN}{Warshall($A,n$)}
1 \'  $W \leftarrow A$\\
2 \' \key{for} \= $k \leftarrow 1$ \key{to} $n$ \\
3 \'           \> \key{do}  \key{for} \= $i \leftarrow 1$ \key{to} $n$ \\
4 \'           \>                      \>  \key{do} \key{for} \= $j \leftarrow 1$ \key{to} $n$ \\ 
5 \'            \>                     \>             \>  \key{do} $w_{ij}\leftarrow w_{ij}+w_{ik}w_{kj}$   \\ 
6 \'  \key{return} $W$
\end{algN}

\noindent From $W$ we obtain easily $R=I+W$.

\noindent For example let us consider the graph in Figure \ref{KZ:fig1}. The corresponding adjacency matrix is:

\[A=\left(\begin{array}{cccccc}
0&0&1&1&1&1 \\
0&0&0&1&1&1 \\
0&0&0&0&1&1 \\
0&0&0&0&0&1 \\
0&0&0&0&0&0 \\
0&0&0&0&0&0 \\
\end{array}
\right) 
\]
After applying the Warshall algorithm we obtain:
\[W=\left(\begin{array}{cccccc}
0&0&1&1&2&3 \\
0&0&0&1&1&2 \\
0&0&0&0&1&1 \\
0&0&0&0&0&1 \\
0&0&0&0&0&0 \\
0&0&0&0&0&0 \\
\end{array}
\right),
\qquad 
R=\left(\begin{array}{cccccc}
1&0&1&1&2&3 \\
0&1&0&1&1&2 \\
0&0&1&0&1&1 \\
0&0&0&1&0&1 \\
0&0&0&0&1&0 \\
0&0&0&0&0&1 \\
\end{array}
\right)
\]
and then $S(6,2)=20,$  the sum of entries in $R$.

The Warshall algorithm combined with the Latin square method can be used to obtain all nontrivial (with length at least 2) super-$d$-subwords of a given  rainbow word $a_1a_2\cdots a_n$. Let us consider a matrix ${\cal A}$ with entries $A_{ij}$ which are set of words. Initially this matrix is defined as:
\[A_{ij}=\left\{  \begin{array}{ll}
                    \{a_ia_j\},  &  \textrm{if } j-i\ge d,\\
                    \emptyset,     &  \textrm{otherwise},
                 \end{array}  \quad \textrm{ for } \, i=1,2,\ldots, n, \, j=1,2,\ldots, n. 
           \right.
\]
If ${\cal A}$ and ${\cal B}$ are sets of words, ${\cal AB}$ will be formed by the set of concatenation of each word from ${\cal A}$ with each word from ${\cal B}$: 
\[ {\cal AB} = \big\{ ab  \, \big| \, a\in {\cal A},  b\in {\cal B}  \big\}.
\] 
If $s=s_1s_2\cdots s_p$ is a word, let us denote by $'s$ the word obtained from $s$ by erasing its first character: $'s=s_2s_3\cdots s_p$.  Let us denote by $'{A_{ij}}$ the set ${A_{ij}}$ in which we erase from each element the first character. In this case $'{\cal A}$ is a matrix with entries $'A_{ij}.$ 

Starting with the matrix ${\cal A}$ defined as before, the algorithm to obtain all nontrivial super-$d$-subwords is the following:

\begin{algN}{Warshall-Latin(${\cal A},n$)}
1 \'  ${\cal W} \leftarrow {\cal A} $\\
2 \' \key{for} \= $k \leftarrow 1$ \key{to} $n$ \\
3 \'           \> \key{do}  \key{for} \= $i \leftarrow 1$ \key{to} $n$ \\
4 \'           \>                      \>  \key{do} \key{for} \= $j \leftarrow 1$ \key{to} $n$ \\ 
5 \'            \>                     \>             \>  \key{do} \key{if} \= $W_{ik}\ne \emptyset$ and  $W_{kj}\ne \emptyset$ \\
6 \'        \>      \>             \>   \> \key{then} $W_{ij}\leftarrow W_{ij} \cup W_{ik}\, 'W_{kj}$   \\ 
7 \'  \key{return} ${\cal W}$
\end{algN}

The set of nontrivial super-$d$-subwords is ${\displaystyle\bigcup_{i,j\in \{ 1,2,\ldots, n \} } W_{ij}}$. 

For $n=8$, $d=3$ the initial matrix is:
\begin{center}
$\left(\begin{array}{cccccccc}
\emptyset & \emptyset & \emptyset & \{ad\} & \{ae\} &\{af\} &\{ag\} &\{ah\} \\ 
\emptyset & \emptyset & \emptyset & \emptyset & \{be\}& \{bf\} & \{bg\}& \{bh\} \\ 
\emptyset & \emptyset & \emptyset & \emptyset & \emptyset & \{cf\}&  \{cg\} & \{ch\} \\ 
\emptyset & \emptyset & \emptyset & \emptyset & \emptyset &\emptyset & \{dg\} &\{dh\} \\ 
\emptyset & \emptyset & \emptyset & \emptyset & \emptyset &\emptyset &\emptyset & \{eh\}\\ 
\emptyset & \emptyset & \emptyset & \emptyset & \emptyset &\emptyset &\emptyset & \emptyset \\
\emptyset & \emptyset & \emptyset & \emptyset & \emptyset &\emptyset &\emptyset & \emptyset \\
\emptyset & \emptyset & \emptyset & \emptyset & \emptyset &\emptyset &\emptyset & \emptyset \\
\end{array}
\right).
$
\end{center}

The result of the algorithm in this case is: 

\begin{center}
$\left(\begin{array}{cccccccc}
\emptyset & \emptyset & \emptyset & \{ad\} & \{ae\} &\{af\} &\{ag, adg\} &\{ah,adh,aeh\} \\ 
\emptyset & \emptyset & \emptyset & \emptyset & \{be\}& \{bf\} & \{bg\}& \{bh,beh\} \\ 
\emptyset & \emptyset & \emptyset & \emptyset & \emptyset & \{cf\}&  \{cg\} & \{ch\} \\ 
\emptyset & \emptyset & \emptyset & \emptyset & \emptyset &\emptyset & \{dg\} &\{dh\} \\ 
\emptyset & \emptyset & \emptyset & \emptyset & \emptyset &\emptyset &\emptyset & \{eh\}\\ 
\emptyset & \emptyset & \emptyset & \emptyset & \emptyset &\emptyset &\emptyset & \emptyset \\
\emptyset & \emptyset & \emptyset & \emptyset & \emptyset &\emptyset &\emptyset & \emptyset \\
\emptyset & \emptyset & \emptyset & \emptyset & \emptyset &\emptyset &\emptyset & \emptyset \\
\end{array}
\right).
$
\end{center}

\section{The general case}
In the general case for any word $w\in \Sigma^*$, let us denote the super-$d$-complexity by $S_w(d)$. We have 
\[ \left\lceil \frac{|w|}{d} \right\rceil  \le S_w(d)\le  S(|w|,d),\]
where $|w|$ is the length of $w$. The minimum value is obtained for a trivial word $w=a\ldots a$, and the maximum one for a rainbow word. 

The algorithm \textsc{Warshall-Latin} can be used for nonrainbow words too, with the remark that repeating subwords must be eliminated. For the word $aabbbaaa$ and $d=3$ the result is: $aa$, $ab$, $aba$, $ba$.

\begin{table}[h]
\begin{center}
\begin{tabular}{|r|rrrrrrrrrr|}\hline
\begin{picture}(15,10)\put(-6,10){\line(2,-1){27}}\put(-2,-2){$n$}\put(13,3){$d$}\end{picture}&  2  & 3  & 4  & 5 &6 &7 &8 &9&10&11\\ \hline
3& 3& -&-&-&-&-&-&-&-&-\\
4& 5&3 &-&-&-&-&-&-&-&-\\
5& 7&5 &3&-&-&-&-&-&-&-\\
6 & 10& 6  &5 &3&-&-&-&-&-&-\\
7 & 14&7  &6 &5 &3 &- &-&-&-&-\\
8 & 19&10  &6& 6 & 5 & 3 &-&-&-&-\\
9 & 26&13  &7 &6 & 6&5&3&-&-&-\\
10 & 35& 15&10 &6 & 6&6&5&3&-&-\\
11 &47& 19&13 &7 & 6&6&6&5&3&-\\
12 &63& 25&14 &10 & 6&6&6&6&5&3\\
\hline
\end{tabular}\caption{Values of $f(2,n,d)$.}
\end{center}
\end{table}

Let us denote by $f(m,n,d)$ the maximum value of the super-$d$-complexity of all words of length $n$ over an alphabet of $m$ letters:

\[ f(m,n,d)= \max_{\scriptsize \begin{array}{c} {w\in \Sigma^n}\\{{ m=|\Sigma|}}\end{array}} \Big(  S_w(d)  \Big). \]

 For $f(2,n,d)$ the following are true, and can be easily proved.

\begin{itemize}\addtolength{\itemsep}{-0.5\baselineskip}
\item $f(2,n,n-1)=3$ for $n\ge 3$.
\item  $f(2,n,n-2)=5$ for $n\ge 4$.
\item If $\displaystyle\left\lceil\frac{n}{2}\right\rceil \le d\le n-3  $ then $f(2,n,d)=6$ for $n\ge 6$.
\item If $n$ is even, then $f\left(2,n,\displaystyle\frac{n-2}{2}\right)=10$ for $n\ge 6$.
\item If $n$ is odd, then $f\left(2,n,\displaystyle\frac{n-1}{2}\right)-7$ for $n\ge 5$.
\end{itemize}

\section*{Conclusions}
The super-$d$-complexity of the finite rainbow words can be obtained by recursive algorithms, by direct mathematical formulas, and by graph algorithms,  all these being presented in this paper. The advantage of the graphs algorithm is that these can be easily altered for obtaining not only the complexity, but the all super-$d$-subwords too. This method can be adapted to obtain the super-$d$-subwords in the general case of the words too, when no restriction on the letters are given.

In the set of all words of a given length over a given alphabet the maximum super-$d$-complexity may be interesting. We present here only some easy to prove results, an extensive study remaining for the future. 

\section*{Acknowledgment}

This work was supported by the project T\'AMOP-4.2.1/B-09/1/KMR-2010-0003 of E\"otv\"os Lor\'and University Budapest.
 
The author is indebted for the useful remarks and suggestions of the unknown referees.


\begin{thebibliography}{99}

\bibitem{KZ:baase} S. Baase, \emph{Computer algorithms: Introduction to design and analysis,} Second edition, Addison--Wesley, 1988.

\bibitem{KZ:CLRS} T. H. Cormen, C. E. Leiserson, R. L. Rivest, C. Stein, \emph{Introduction to algorithms}, Second edition, The MIT Press/McGraw Hill, Cambridge/Boston, 2001. 

\bibitem{KZ:fiz} W. Ebeling, R. Feistel, \emph{Physik der Selbstorganisation und Evolution,} Akademie-Verlag, Berlin, 1982. 

\bibitem{KZ:elzinga} C. Elzinga, S. Rahmann, H. Wung, Algorithms for subsequence combinatorics, \emph{Theor. Comput. Sci.}, \textbf{409,} 3 (2008) 394--404.

\bibitem{KZ:elzinga2} C. H. Elzinga,  Complexity of categorial time series,  \textit{Sociological Methods \emph{\&} Research},  \textbf{38,} 3 (2010) 463--481. 

\bibitem{KZ:ivanyi} A. Iv\'anyi, On the $d$-complexity of words, \emph{Annales Univ. Sci. Budapest., Sect. Computatorica,}
\textbf{8} (1987) 69--90.  

\bibitem{KZ:kasa} Z. K\'asa, On the $d$-complexity of strings, \emph{Pure Math. Appl.}, \textbf{9,} 1--2 (1998) 119--128.

\bibitem{KZ:online} N. J. A. Sloane,   The on-line encyclopedia of integer sequences,\\  
\texttt{http://www.research.att.com/\~{}njas/sequences/}.

\bibitem{KZ:bio} O. G. Troyanskaya, O. Arbell, Y. Koren, G. M. Landau, A. Bolshoy, Sequence complexity profiles of prokaryotic genomic sequences: A fast algorithm for calculating linguistic complexity, \emph{Bioinformatics,} \textbf{18,} 5 (2002) 679--688.

\end{thebibliography}
\end{document}